\documentclass[twoside,12pt]{article}
\usepackage{epsfig}

\def\Journal#1#2#3#4{{#1} {#2} (#4) #3 }

\def\NPA{{\em Nucl. Phys.} A}

\def\PLB{{\em Phys. Lett.} B}

\def\PRD{{\em Phys. Rev.} D}
\def\PRC{{\em Phys. Rev.} C}

\def\RMP{{\em Rev. Mod. Phys.}}

\def\noin{\noindent}

\newcommand{\be}{\begin{equation}}
\newcommand{\ee}{\end{equation}}
\newcommand{\bea}{\begin{eqnarray}}
\newcommand{\eea}{\end{eqnarray}}

\begin{document}

\title{ \vspace{1cm} Effective Field Theory and Electroweak Processes in Nuclei}
\author{K.\ Kubodera\\ 
\\
Department of Physics \& Astronomy, University of South Carolina, Columbia, USA}
\maketitle
\begin{abstract} 
The accurate theoretical treatment of low-energy electro-weak processes
in lightest nuclei is of great current importance not only in the context of
nuclear physics {\it per se} but also from the astrophysical and particle-physics
viewpoints.  I give a brief survey of the recent remarkable progress in this domain.

\end{abstract}
%\eject
%\tableofcontents

\noin
Rigorous theoretical treatments of low-energy electroweak processes 
in lightest nuclei are important for multiple reasons.
First, the accurate understanding of these processes is a step 
required before we can hope to get a reliable framework for describing 
electroweak processes in heavier nuclear systems.
Secondly, many of these reactions
feature importantly in astrophysical processes,
and/or in neutrino physics experiments.
This second aspect is getting ever increasing attention,
and the purpose of this short note is to describe some
of the notable developments that have occurred on this topic
in the recent years.
I shall be concerned with the following reactions:
\noin
(i) $pp$-fusion:
$p\!+\!p\to d\!+\!e^+\!\!+\! \bar{\nu_e}$;
(ii) $\nu$-$d$ reactions:
$\nu_e\!+\! d \to e^-\!\!+\!p\!+\!p$,\,\,\,
$\nu\!+\!d \to \nu\!+\!n\!+\!p$;  
(iii) $\mu$-$d$ capture:
$\mu^-\!+\! d \to \nu_e\!+\!n\!+\!n$;
(iv) Hep and Hen:
$^3{\rm He}\!+\!p \to ^4\!\!{\rm He}\!+\!\nu_e\!+\! e $,\,
$^3{\rm He}\!+\!n \to ^4\!{\rm He}\!+\!\gamma $;
(v) radiatice pion capture:
$\pi^- d \to \gamma nn$. 
This list may look like a random collection of unrelated reactions,
but I hope to show they are all related to the goal 
of improving the precision of theoretical predictions 
for low-energy electroweak processes in lightest nuclei. 

To present my point succinctly, let me concentrate on $pp$-fusion 
and consider the $pp$-fusion $S$-factor, $S_{pp}$;
this quantity is obtained after removing
from the $pp$-fusion cross section the ``trivial" energy dependence
such as the $1/E$ dependence and the Coulomb penetrability.
The solar model and stellar evolution theory have now reached 
a level that  requires $\sim$1\% precision in $S_{pp}$,
whereas the theoretical uncertainty in $S_{pp}$
quoted in the celebrated 1998 review article~\cite{ADEetal98}
is as large as $\sim$6 \%.
How to go beyond this level was an urgent question.
Recent theoretical investigations based on the nuclear physics
applications of effective field theory (EFT) have significantly 
improved the situation, as will be described below.

Before going into this main topic,
let us take a quick look at the present situation 
of relevant lattice QCD calculations. 
The latest lattice QCD calculation of the nucleon weak form factors
has been reported in Ref.~\cite{Lattice2}. 
Just to illustrate the level of precision achieved in this highly 
elaborate work, I mention that the calculated value of 
$g_A/g_V=1.19 (6)_{\rm stat}(4)_{\rm syst}$
is 7\% smaller than the experimental value.
So, although it is obviously very important to further push
{\it ab ititio} calculations based on lattice QCD,
there seems to be a while before
one can reach 1\% precision in calculating 
the electroweak properties of the nucleon,
and the challenge should be even harder for the two-nucleon 
systems.
This situation enhances the relevance of EFT-based studies.

\noin
{\bf Effective field theory (EFT)} -----
The basic idea of EFT is that,
to describe low energy-momentum phenomena
characterized by a scale $Q$,
we can introduce a cut-off scale $\Lambda_{\rm cut}\gg Q$
and choose to retain only low energy-momentum degrees of freedom 
(effective fields $\phi_{\rm eff}$).
The effective Lagrangian ${\cal L}_{\rm eff}$
consists of those monomials of $\phi_{\rm eff}$ and its derivatives
which are consistent with the symmetries.
Since a term involving $n$ derivatives scales like 
$(Q/\Lambda_{\rm cut})^n$,
we have a perturbative series in $Q/\Lambda_{\rm cut}$.
The coefficient of each term, 
called the low-energy constant (LEC),
subsumes the high-energy physics that has been ``integrated away".
If all the LEC's up to a specified order $n$ are known,
${\cal L}_{\rm eff}$ serves as a complete 
(and hence model-independent) Lagrangian.
In the nuclear physics application of EFT,
the underlying Lagrangian is that of QCD, 
$\phi_{\rm eff}$ represents the nucleons and pions,
with $\Lambda_{\rm cut}\sim$ 1 GeV;
the corresponding EFT is known as chiral perturbation theory ($\chi$PT).
$\chi$PT cannot be applied 
in a straightforward manner to nuclei because of 
the existence of very low-lying excited states in nuclei.
Weinberg~\cite{wei90} proposed to resolve this difficulty
by classifying Feynman diagrams into irreducible and reducible
diagrams and apply the chiral counting rules
only to irreducible diagrams.
Treating irreducible diagrams 
(up to a specified chiral order)
as an effective potential
(to be denoted by $V_{ij}^{\mbox{\tiny{EFT}}}$,
$V_{ijk}^{\mbox{\tiny{EFT}}}$, etc.)
acting on nuclear wave functions,
one can incorporate the reducible diagrams
by solving the Schr\"odinger equation
$H^{\mbox{\tiny{EFT}}}|\Psi^{\mbox{\tiny{EFT}}}\!\!>\,=\,
\!E|\Psi^{\mbox{\tiny{EFT}}}\!\!>$ 
with 
$H^{\mbox{\tiny{EFT}}}\,=\,
\sum_i^A t_i + \sum_{i<j}^A V_{ij}^{\mbox{\tiny{EFT}}}
+\sum_{i<j<k}^A V_{ijk}^{\mbox{\tiny{EFT}}}$.
An electroweak transition matrix in nuclear EFT
is given by 
${\cal M}_{fi}^{\mbox{\tiny{EFT}}}\!=
<\!\!\Psi_{\!f}^{\mbox{\tiny{EFT}}}\,|
{\cal T}^{\mbox{\tiny{EFT}}}
\,|\Psi_i^{\mbox{\tiny{EFT}}}\!>
= <\!\!\Psi_f^{\mbox{\tiny{EFT}}}|
\sum_i^A{\cal O}_i^{\mbox{\tiny{EFT}}}
\!\!+\!\sum_{i<j}^{A}{\cal O}_{ij}^{\mbox{\tiny{EFT}}}\!+\!\,\cdots
|\Psi_i^{\mbox{\tiny{EFT}}}\!\!>\,,$
where $ {\cal T}^{\mbox{\tiny{EFT}}}
\!= \!\sum_i^A{\cal O}_i^{\mbox{\tiny{EFT}}}
+\sum_{i<j}^{A}{\cal O}_{ij}^{\mbox{\tiny{EFT}}} +\, \cdots \,$.
is the relevant transition operator  
derived in EFT up to a specified chiral order.
It is however a non-trivial task to fully carry out this program
because of difficulties in getting $\Psi^{\rm EFT}$,
and because determining all the LECs involved
can be challenging.
To cope with these problems, Park {\it et al.}~\cite{PARetal03}
introduced a hybrid EFT approach called EFT*
in which $\Psi^{\mbox{\tiny{EFT}}}$ is replaced 
with $\Psi^{\mbox{\tiny{phen}}}$ that has been obtained
from the high-precision phenomenological $NN$ potential.
A notable merit of EFT* is that it is applicable to complex nuclei
(A = 3,4 \ldots) with essentially the same accuracy and ease
as to the A=2 system.
This opens up the possibility to determine LEC(s)
using observables pertaining to complex nuclei.
Mismatch between $\Psi^{\mbox{\tiny{EFT}}}$ and 
$\Psi^{\mbox{\tiny{phen}}}$ is expected to affect
only short-distance behavior.
If one introduces a momentum cutoff parameter
$\Lambda_{\rm NN}$ to tame the short-distance behavior,
the degree of mismatch is likely 
to be reflected in the sensitivity of the result
to $\Lambda_{\rm NN}$.
Converesly, $\Lambda_{\rm NN}$-independence
may be taken as a measure of model independence
of an EFT* calculation.

Park {\it et al.}~\cite{PARetal03} carried out an EFT* calculation of the 
Gamow-Teller (GT) transitions in the A=2, 3 and 4 systems.
A crucial point here is that $pp$-fusion, $\nu$-$d$ reactions, 
$\mu$-$d$ capture, tritium $\beta$-decay and Hep are all
controlled by the single common LEC, $\widehat{d}^R$,
which is the strength of contact-type
four-nucleon coupling to the axial current.
From the the tritium $\beta$-decay rate $\Gamma_{\beta}^t$
known with high precision, one can determine $\widehat{d}^R$
and subsequently carry out EFT* calculations of
$pp$-fusion, $\nu d$ scattering and $\mu d$ capture.
A next-to-next-to-leading order (NNLO) calculation
of $pp$-fusion in Ref.~\cite{PARetal03} 
yields $S_{pp}=  3.94\times(1\pm0.004)\times10^{-24}$ MeV b.
The uncertainty due to the $\Lambda_{\rm NN}$ dependence
is found to be much smaller than
the uncertainty caused by the experimental error   
in $\Gamma_\beta^t$.
For other EFT-based calculations of $S_{pp}$,
see Refs.~\cite{KR01,ANDetal08a}.

With the use of the same $\widehat{d}^R$,
Ando {\it et al.}~\cite{Aetal03} made an EFT* calculattion
of the $\nu$-$d$ cross sections; see also Ref.~\cite{BCK01}.
These cross sections are of great importance 
in connection with the SNO experiments
and have been studied intensively for many years;
see {\it e.g.,} Ref.~\cite{KN}.
The results in Ref.~\cite{Aetal03} agree 
with those obtained by 
Nakamura {\it et al.}~\cite{NSGK,Netal} 
with the use of the so-called 
standard nuclear physics approach (SNPA),
but the EFT* calculation has significantly
reduced the theoretical errors. 

The availability of a reliable estimate of $\widehat{d}^R$
also prompted an EFT* calculation of the
$\mu d$ capture rate, $\Gamma_{\mu d}$,
by Ando {\it et al.}~\cite{APKM02};
see also Ref.~\cite{CHEetal05}.
In this connection, it is to be noted that
the MuSun Collaboration~\cite{MuSun}
is aiming to measure $\Gamma_{\mu d}$
with $\sim$1 \% precision.
The result of this experiment 
will allow us to determine $\widehat{d}^R$
using an observable in the A=2 system.
Although there is a reason to believe that
the $\widehat{d}^R$ determined 
from $\Gamma_{\beta}^t$ is reliable, 
it is certainly nice if we can fix
$\hat{d}^R$ from an A=2 observable
such as $\Gamma_{\mu d}$. 

As for the reliability of the EFT* formalism itself, 
Lazauskas {\it et al.}'s recent work 
on Hen~\cite{LAZetal09} is highly informative.
The Hen reaction shares with Hep  
the unique feature that
the formally leading-order one-body term 
is highly suppressed and that 
destructive interference 
between the suppressed one-body term and the two-body term
leads to a further drastic reduction of the cross section.
This feature renders the calculation of Hep and Hen
quite non-trivial, but it also implies that 
these reactions are good places for studying
higher chiral order contributions.
An N$^3$LO calculation in Ref.~\cite{LAZetal09}
gives $\sigma_{\rm Hen}= (38 \sim 58)$ $\mu b$,
with high stability against changes in 
the cutoff  parameter $\Lambda_{NN}$.
The calculated value agrees with  
$\sigma_{\rm Hen}^{\rm exp}=
(54\pm6)$ $\mu b$.  
The success of EFT* for Hen
not only establishes the reliability 
of  the earlier EFT* calculation of Hep~\cite{PARetal03}
but also provides strong support for the EFT* framework
in general.

I now briefly discuss other recent developments 
that contribute to the improvement of precision in $S_{pp}$.
The current experimental error 
in the neutron-neutron scattering length, $a_{nn}$, 
is estimated to cause $0.5\%$ error in $S_{pp}$.
It has been pointed out 
(see, {\it e.g.,} Ref.~\cite{GAR09})
that the use of EFT in analyzing 
the $\pi^- d \to \gamma nn$ reaction can 
significantly reduce the errors in the value of $a_{nn}$.
When we aim at 1\% precision in $S_{pp}$,
radiative corrections play an important role.
A recent detailed study~\cite{KURetal03}
indicate that, with the use of the Fermi constant
$G_F$ obtained from $\mu$-decay
and the ``effective" $g_A$ obtained 
from neutron $\beta$-decay,
radiative corrections specific to $pp$-fusion 
are $\sim$ 3-4 \% effects, and these can be 
estimated within 0.1 \% uncertainty.
See also Ref.~\cite{FK05}.

To summarize, 
$S_{pp}$ can be calculated with $\sim$ 1\% precision.
The results are (in all likelihood) already available 
from the combination SNPA and EFT*.
The MuSun experiment is very important in that
it will enable us to determine $\widehat{d}^R$
within the A=2 system. 
%(without resorting to tritium $\beta$-decay).
Full EFT calculations
that use $\Psi^{\rm EFT}$ instead of
$\Psi^{\rm phen}$ are eagerly awaited.
For the construction
of EFT-based nuclear interactions,
see {\it e.g.,} Ref.~\cite{Epelbaum06}.

\vspace{2mm}
The present work is supported in part by the NSF Grant No.~PHY-0758114.


\begin{thebibliography}{99}
\itemsep -2pt 

\bibitem{ADEetal98} E.G. Adelberger {\it et al.}, 
\Journal{\RMP} {70} {1265} {1998}

\bibitem{Lattice2}
T. Yamazaki et al. (RBC and UKQCD Collaborations),
\Journal{\PRD} {79} {114505} {2009}

\bibitem{wei90}
S. Weinberg, \Journal{\PLB} {251} {288} {1990};
\Journal{\PLB} {295} {114} {1992}

\bibitem{PARetal03}
T.-S. Park et al.
\Journal{\PRC} {67} {055206} {2003}

\bibitem{KR01}
X. Kong and F. Ravndal,
\Journal{\PRC} {64} {044002} {2001}


\bibitem{ANDetal08a}
S. Ando, J.~W. Shin, C.~H. Hyun, S.~W. Hong and
K. Kubodera,
\Journal{\PLB}{668}{187} {2008}

\bibitem{Aetal03} 
S. Ando {\it et al.,} 
\Journal{\PLB} {555} {49} {2003}

\bibitem{BCK01}
M. Butler, J.-W. Chen and X. Kong, 
\Journal{\PRC}{63} {035501} {2001}

\bibitem{KN}
K. Kubodera and S. Nozawa,
{\em Int. J. Mod. Phys.} E 3 (1994) 101

\bibitem{NSGK}
S. Nakamura, T. Sato, V. Gudkov and K. Kubodera,
\Journal{\PRC} {63} {034617} {2001}

\bibitem{Netal} 
S. Nakamura {\it et al.,} 
\Journal{\NPA} {707} {561} {2002}


\bibitem{APKM02} 
S. Ando, T.-S. Park, K. Kubodera and F. Myhrer,
\Journal{\PLB} {533} {25} {2002}

\bibitem{CHEetal05}
J.-W. Chen, T. Inoue, X.-D. Ji and Y.-C. Li,
\Journal{\PRC} {72} {061001} {2005}

\bibitem{MuSun}
V.~A. Andreev et al. (MuSun Collaboration),\\
URL {\tt http://www.npl.uiuc.edu/exp/musun} (2008)

\bibitem{LAZetal09}
R. Lazauskas, Y.-H. Song and T.-S. Park, arXiv: 0905.3119 [nucl-th]

\bibitem{GAR09}
A. Gardstig, {\em J. Phys.} G 36 (2009) 053001 

\bibitem{KURetal03}
A. Kurylov, M.~J. Ramsey-Musolf and P. Vogel,
\Journal{\PRC} {67} {035502}{2003}

\bibitem{FK05}
M. Fukugita and T. Kubota,
\Journal{\PRD} {72} {071301} {2005}

\bibitem{Epelbaum06}
E. Epelbaum, 
{\em Prog. Part. Nucl. Phys.}  57 (2006) 654





\end{thebibliography}
\end{document}